\documentclass[acmsmall,screen]{acmart}
\AtBeginDocument{%
  }

\setcopyright{acmlicensed}
\copyrightyear{2026}
\acmYear{2026}
\acmDOI{XXXXXXX.XXXXXXX}

\acmJournal{tosem}

\usepackage[framemethod=tikz]{mdframed}
\newcounter{finding}

\usepackage{tcolorbox,tabularx,graphicx,xcolor,tikz,makecell,enumitem,longtable,booktabs}
\usetikzlibrary{decorations.pathmorphing}
\usepackage[edges]{forest}

\begin{document}

\title{Promptware Engineering: Software Engineering for Prompt-Enabled Systems}

\author{Zhenpeng Chen}
\email{zpchen@tsinghua.edu.cn}
\affiliation{%
  \institution{Tsinghua University}
  \country{China}
}

\author{Chong Wang}
\email{chong.wang@ntu.edu.sg}
\affiliation{%
  \institution{Nanyang Technological University}
  \country{Singapore}
}\authornote{Corresponding author.}

\author{Weisong Sun}
\email{weisong.sun@ntu.edu.sg}
\affiliation{%
  \institution{Nanyang Technological University}
  \country{Singapore}
}

\author{Xuanzhe Liu}
\email{liuxuanzhe@pku.edu.cn}
\affiliation{%
  \institution{Peking University}
  \country{China}
}

\author{Jie M. Zhang}
\email{jie.zhang@kcl.ac.uk}
\affiliation{%
  \institution{King's College London}
  \country{United Kingdom}
}

\author{Yang Liu}
\email{yangliu@ntu.edu.sg}
\affiliation{%
  \institution{Nanyang Technological University}
  \country{Singapore}
}


\renewcommand{\shortauthors}{Chen et al.}

\begin{abstract}
Large Language Models (LLMs) are increasingly integrated into software applications, giving rise to a broad class of prompt-enabled systems, in which prompts serve as the primary `programming' interface for guiding system behavior. Building on this trend, a new software paradigm, \emph{promptware}, has emerged, which treats natural language prompts as first-class software artifacts for interacting with LLMs. Unlike traditional software, which relies on formal programming languages and deterministic runtime environments, promptware is based on ambiguous, unstructured, and context-dependent natural language and operates on LLMs as runtime environments, which are probabilistic and non-deterministic. These fundamental differences introduce unique challenges in prompt development. In practice, prompt development remains largely ad hoc and relies heavily on time-consuming trial-and-error, a challenge we term the \emph{promptware crisis}. To address this, we propose \emph{promptware engineering}, a new methodology that adapts established Software Engineering (SE) principles to prompt development. Drawing on decades of success in traditional SE, we envision a systematic framework encompassing prompt requirements engineering, design, implementation, testing, debugging, evolution, deployment, and monitoring. Our framework re-contextualizes emerging prompt-related challenges within the SE lifecycle, providing principled guidance beyond ad-hoc practices. Without the SE discipline, prompt development is likely to remain mired in trial-and-error. This paper outlines a comprehensive roadmap for promptware engineering, identifying key research directions and offering actionable insights to advance the development of prompt-enabled systems.
\end{abstract}

\ccsdesc[500]{Software and its engineering~Software creation and management}
\ccsdesc[500]{Computing methodologies~Machine learning}

\keywords{Promptware Engineering, Software Engineering, LLM, Prompt}

\received{2025}
\received[revised]{2026}
\received[accepted]{2026}

\settopmatter{printfolios=true} 
\maketitle

\section{Introduction}
Large Language Models (LLMs), such as GPT~\cite{achiam2023gpt}, LLaMA~\cite{touvron2023llama}, and DeepSeek~\cite{bi2024deepseek}, are increasingly integrated into software applications across diverse domains~\cite{xiao2025bias,wang2024large,li2023large,guo2026eet}. Major technology companies, including Microsoft, Google, Amazon, and Apple, have integrated LLMs into their software products \cite{nahar2024beyond}, reaching millions of users.

This trend has given rise to a broad class of prompt-enabled systems, in which LLMs are embedded as core components and prompts serve as the primary `programming' interface for shaping the behavior and outputs of LLMs~\cite{tafreshipour2024prompting}. Recognizing the central role of prompts, practitioners have introduced template-based approaches, such as the Liquid prompt template~\cite{liquiddoc} and the LangChain prompt template~\cite{oshin2025learning}, to support prompt programming.
Building on these developments, \emph{promptware} has emerged as a new software paradigm that treats natural language prompts as first-class software artifacts, complementing or encapsulating traditional code to enable software functionality through direct interaction with LLMs~\cite{HassanLRGC00TOL24}.

Promptware fundamentally differs from traditional software in two key aspects: language and runtime environment. Unlike structured programming languages with strict syntax and deterministic behavior, prompts are written in natural language, which is flexible, context-dependent, and ambiguous. This makes formalizing prompt construction and analysis challenging. Additionally, while traditional software typically relies on deterministic runtime environments, promptware uses probabilistic, non-deterministic LLMs as runtime environments, introducing unique challenges such as human-like behaviors, unclear capability boundaries, undefined error handling, and uncertain execution control. These complexities make prompt development particularly difficult.

Prompt engineering has been a widely adopted solution for prompt development~\cite{sahoo2024systematic}. OpenAI defines it as `designing and optimizing input prompts to effectively guide a language model's responses'~\cite{openaiPEdefine}; Meta calls it `a technique used in natural language processing to improve the performance of the language model by providing more context and information about the task at hand'~\cite{metaPEdefine}. These organizations have published official guidelines to support prompt engineering~\cite{openaiguide, metaPEdefine}, but they share a critical limitation: they are output-centric and lack systematic methodologies for prompt development.

In practice, prompt engineering often relies on ad hoc, experimental approaches~\cite{LLMcain2025, tafreshipour2024prompting, PromptsArePrograms, nahar2024beyond, parnin2023building, hassan2024rethinking, dolata2024development}, especially as prompt-enabled systems become increasingly complex—what we refer to as the \emph{promptware crisis}. Existing studies~\cite{dolata2024development,parnin2023building,PromptsArePrograms,LLMcain2025} show that prompt engineering is largely trial-and-error, time-consuming, and challenging, with even experienced engineers facing difficulties. This highlights the urgent need for a systematic, development-centric framework to move beyond the ad hoc approaches and address the growing challenges of the promptware crisis.

In this paper, we introduce \emph{promptware engineering}, a new methodology that applies Software Engineering (SE) principles to prompt development, providing a unified and principled framework to address prompt-related challenges systematically.
This vision is driven by two key rationales: (1)~As LLMs become integral to an expanding range of software applications, prompts have emerged as crucial software components. Thus, promptware has surfaced as an evolving software paradigm, with prompt development increasingly recognized as a new form of programming~\cite{PromptsArePrograms}. This shift positions prompt development as a vital aspect of SE. (2)~Current prompt development practices are often ad hoc and experimental, driven by unique complexities. However, decades of SE advancements highlight the value of systematic approaches to manage complexity, ensure quality, and support iterative improvements, underscoring the potential to transform prompt development from an experimental practice into a structured and disciplined process.

Realizing the vision of promptware engineering requires integrating core SE activities, specifically tailored to the unique demands of prompt development. These activities include prompt requirements engineering, design, implementation, testing, debugging, evolution, deployment, and monitoring. To support them, we emphasize the need for innovative research, tools, and automation throughout the prompt development lifecycle. This paper discusses key challenges and highlights promising research opportunities in promptware engineering.

\section{Preliminaries}
This section defines key terminologies and situates our vision within the context of existing research.

\subsection{Terminologies}
LLMs, such as GPT~\cite{achiam2023gpt} and LLaMA~\cite{touvron2023llama}, are increasingly integrated into software applications to perform diverse tasks~\cite{LLMcain2025,liu2025first,yang2025bamas}. Such LLM-based software applications are also referred to as \emph{prompt-enabled systems}~\cite{PromptsArePrograms}, in which prompts serve as the primary interface for interacting with LLMs. \emph{Prompts} are natural language instructions or queries that define the context, task, or expected behavior of LLMs~\cite{tafreshipour2024prompting}, enabling these systems to operate without modifying model parameters. \emph{Promptware}~\cite{HassanLRGC00TOL24} denotes the software paradigm in which natural language prompts are used as core artifacts to interact with LLMs, while \emph{promptware engineering} systematically applies SE principles to prompt development. 

Rather than replacing traditional SE, promptware engineering represents its natural evolution in the era of LLMs, where established SE activities such as requirements engineering, design, testing, and deployment remain essential but must be reinterpreted for a setting characterized by natural-language-based programming interfaces, probabilistic and non-deterministic execution, dynamically evolving models, and human-like reasoning that introduces ambiguity and bias.

\subsection{Related Work}
There have been numerous studies combining prompt engineering and SE.
Most of these works focus on prompt engineering for SE. The widespread adoption of LLMs in SE tasks has sparked significant interest in prompt engineering techniques aimed at improving the effectiveness of LLMs in these scenarios. Alshahwan et al.~\cite{AlshahwanHHMSW24} proposed a vision for assured LLM-based SE, where prompt engineering is leveraged as a component of a search-based SE process. Wang et al.~\cite{wang2024software} conducted a survey on the use of LLMs in software testing, a critical SE activity. Their findings revealed that about two-thirds of studies in this domain employ prompt engineering.
Fan et al.~\cite{FanGHLSYZ23} and Hou et al.~\cite{hou2024large} extended the scope of the surveys to encompass the entire SE lifecycle. Fan et al.~\cite{FanGHLSYZ23} reviewed common prompt engineering strategies applied to tasks such as code generation, software testing, and maintenance. Hou et al.~\cite{hou2024large} identified eight prompt engineering techniques currently employed in the LLM for SE domain. 
Recently, the rise of LLM-based agents in SE has further highlighted the importance of prompt engineering techniques. These techniques have been widely adopted in constructing agents to enhance their performance across various SE tasks~\cite{liu2024large}. 

In contrast, SE for prompt engineering has not been well explored. Existing studies primarily focus on empirical investigations into the challenges software engineers face during prompt engineering.
Mailach et al.~\cite{LLMcain2025} and Parnin et al.~\cite{parnin2023building} studied challenges faced by software practitioners in developing LLM-based applications, and identified prompt engineering as a key concern. Practitioners highlighted issues such as the experimental nature of prompting, as well as challenges related to context length, computational costs, and unpredictable changes in prompts~\cite{LLMcain2025}.
They also noted that designing and managing prompts efficiently is time-consuming and resource-constrained~\cite{parnin2023building}.
Similarly, Dolata et al.~\cite{dolata2024development} found that freelancers struggle with trial-and-error prompting cycles. They reported that subtle differences in prompts resulted in inconsistent outputs and identified challenges in managing experimentation costs.
Liang et al.~\cite{PromptsArePrograms} conceptualized prompts as a new type of program and conducted interviews to understand how developers integrate them into software. Their findings revealed that prompt programming is a rapid, unsystematic process, significantly different from traditional software development.
Tafreshipour et al.~\cite{tafreshipour2024prompting} studied prompt evolution in real-world projects and reported that only 21.9\% of prompt changes are documented. These changes often result in logical inconsistencies and misalignments between prompts and LLM responses.
Nahar et al.~\cite{nahar2024beyond} interviewed product teams at Microsoft and confirmed numerous challenges in integrating LLMs into software products, including those specifically related to prompt engineering. These studies emphasize the pressing need to transition prompt development from the current experimental approach to a systematic framework.

Recently, Hassan et al.~\cite{HassanLRGC00TOL24} revisited SE in the foundation model era and identified crafting effective prompts as an emerging challenge. They proposed solutions such as prompt-aware IDEs to assist developers in crafting effective prompts. In follow-up work~\cite{hassan2024rethinking}, they further envisioned prompt transpilers to reduce the burden of prompt crafting, shifting responsibility from humans to AI systems. While these studies provide valuable insight into prompt-related challenges and potential tooling, they primarily focus on prompt development as a single step rather than treating prompts as full-fledged software artifacts. In contrast, promptware engineering adopts a full-lifecycle, engineering-centric perspective, systematically applying SE principles across the prompt lifecycle. By framing prompts as software artifacts with explicit lifecycle considerations, our framework moves beyond isolated tools or challenge identification and provides structured, systematic methods for building reliable and maintainable prompt-based systems.

\begin{figure*}[t]
    \centering
\includegraphics[width=1\linewidth]{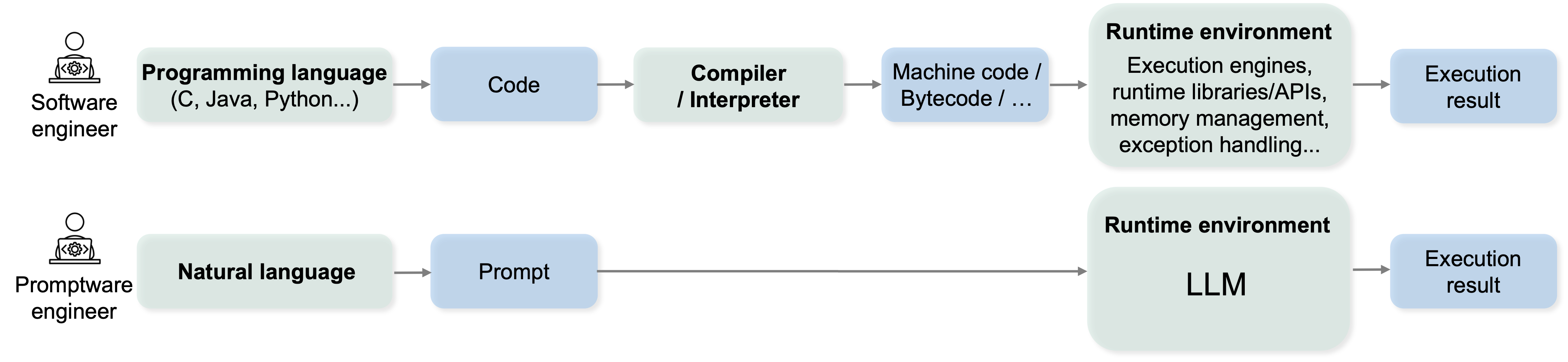}
  \caption{Comparison of the traditional software paradigm and promptware.}
  \label{fig:diff} 
\end{figure*}

\section{Traditional Software Paradigm vs. Promptware}
Before exploring the research opportunities in promptware engineering, we first compare promptware with the traditional software paradigm to analyze its unique characteristics.

Figure~\ref{fig:diff} illustrates the comparison.
Traditionally, software engineers write code using programming languages such as C, Java, and Python. Depending on the language, a compiler or interpreter translates the code into machine code, bytecode, or an intermediate representation. The translated code is then executed within a runtime environment, which typically includes components such as execution engines, runtime libraries, memory management systems, and exception handling mechanisms.

In contrast, promptware engineers write prompts primarily using natural language, eliminating the need for traditional compilation or interpretation. Instead, an LLM serves as the runtime environment, interpreting the input prompt and generating responses based on probabilistic reasoning and pre-trained knowledge, often in real-time.

This comparison indicates that the unique characteristics of promptware primarily arise from the distinct nature of its \emph{language} and \emph{runtime environment}. In the following, we outline these specific characteristics, labeled as C1, C2, …, C10. A summary of the comparison between promptware and the traditional software paradigm in terms of language is presented in Table~\ref{lang_diff}, while the comparison regarding the runtime environment is summarized in Table~\ref{run_diff}.

\begin{table*}[!tp]
\footnotesize
\centering
\caption{Traditional programming language vs. natural language.}
\label{lang_diff}
\begin{tabularx}{\linewidth}{p{2.2cm}p{4.8cm}p{6cm}}
\hline
\textbf{Aspect} & \textbf{Programming Language} & \textbf{Natural Language}\\
\hline
\textbf{C1. Structure and Formalism} & Highly structured with rigorous syntax and well-defined semantics. & Unstructured, flexible, and open-ended.\\
\hline
\textbf{C2. Explicitness and Determinism} & Explicit;  deterministic behavior. & Ambiguous and context-dependent; probabilistic behavior.\\
\hline
\textbf{C3. Correctness and Quality} & Explicit correctness defined by formal specifications; syntax errors are fatal. & No universal correctness; typos or grammatical errors may not cause failures, but can subtly alter meaning.\\
\hline
\end{tabularx}
\end{table*}

\begin{table*}[!tp]
\footnotesize
\centering
\caption{Traditional runtime environment  vs. LLM as runtime environment.}
\label{run_diff}
\begin{tabularx}{\linewidth}{p{2.2cm}p{4.9cm}p{5.9cm}}
\hline
\textbf{Aspect} &  \textbf{Traditional Runtime Environment} & \textbf{LLM-as-Runtime Environment}\\
\hline
\textbf{C4. Determinism} & Same input typically yields the same output. & Same prompt can yield different outputs.\\
\hline
\textbf{C5. Human-Like Characteristics} & Executes code mechanically following predefined logic. & Exhibits human-like characteristics (e.g., contextual reasoning and social intelligence).\\
\hline
\textbf{C6. Capability Boundaries} & Well-defined capability boundaries; engineers understand execution rules. & Unclear, evolving capability boundaries; execution functions as a black box.\\
\hline
\textbf{C7. Error Handling} & Deterministic error handling, with clear error messages, stack traces, and debugging mechanisms. & Implicit and unpredictable fault tolerance, without explicit error signals.\\
\hline
\textbf{C8. Execution Control} & Precise execution control: step through code, inspect variables, use debuggers. & Indirect control: adjust prompts heuristically experimentally.\\
\hline
\textbf{C9. Access Control} & Strict access control: permissions, memory protections, sandboxing. & Limited access control: vulnerable to security risks, lack of rigid execution boundaries.\\
\hline
\textbf{C10. Memory Management} & Explicit memory management: direct control over allocation and deallocation. & No persistent memory: contextual processing without long-term memory unless external memory mechanisms are used.\\
\hline
\end{tabularx}
\end{table*}

\subsection{Traditional Programming Language vs. Natural Language}
\noindent \textbf{C1. Structure and Formalism:} Traditional programming languages are highly structured, with rigorous syntax and well-defined semantics that enforce precise coding rules. In contrast, natural language is inherently unstructured, flexible, and open-ended, making it difficult to establish formalized rules for prompt construction.

\noindent \textbf{C2. Explicitness and Determinism:} Traditional programming languages are explicit and deterministic, meaning the same code typically produces the same output. This determinism enables well-established techniques for analyzing aspects such as data flow, control flow, and dependencies. In contrast, natural language is ambiguous and context-dependent, relying on prior knowledge, pragmatics, and contextual cues rather than rigid rules. Its probabilistic nature makes it difficult to ensure consistent and reliable prompt outcomes, as well as to systematically analyze how prompts influence model outputs.

\noindent \textbf{C3. Correctness and Quality:} In traditional programming, correctness is explicitly defined through formal specifications, and both functional and non-functional quality attributes can be systematically measured. Even minor syntax errors can lead to failures or bugs. In contrast, natural language lacks a universal standard for correctness but follows conventions and guidelines (such as grammar rules and clarity of intent) that affect a prompt’s effectiveness. While minor typos or grammatical inconsistencies may not always cause immediate failures, they can subtly alter meaning, potentially leading to unpredictable or undesired outcomes.

\subsection{Traditional Runtime Environment vs. LLM-as-Runtime Environment}
\noindent \textbf{C4. Determinism:} Traditional runtime environments are deterministic, meaning the same input typically produces the same output under identical conditions. Execution follows well-defined rules dictated by the programming language, system architecture, and compiler or interpreter behavior. In contrast, LLMs generate responses probabilistically based on learned statistical patterns, resulting in inherent non-determinism. The same prompt can produce different outputs across executions, posing challenges for reproducibility, reliability, and consistency in promptware engineering.

\noindent \textbf{C5. Human-Like Characteristics:} Traditional runtime environments execute code mechanically without subjective interpretation. In contrast, LLMs exhibit human-like characteristics, such as contextual reasoning, emotional alignment, and social intelligence. While these traits enhance flexibility, they introduce challenges in control, intent alignment, and reliability. LLMs may generate responses influenced by biases, emotional tone, or implicit assumptions, making promptware harder to stabilize, predict, and interpret.
 
\noindent \textbf{C6. Capability Boundaries:} Traditional runtime environments have well-defined capability boundaries, with execution governed by explicit rules and formal models. Engineers can analyze and predict system behavior based on well-documented specifications. In contrast, LLMs have unclear and evolving capability boundaries, often exhibiting emergent behaviors. Moreover, LLM execution functions as a black box—developers lack direct visibility into how prompts are internally processed, making it difficult to predict, control, or explain outcomes.

\noindent \textbf{C7. Error Handling:} Traditional runtime environments handle errors deterministically, providing well-defined error messages and stack traces. Errors are explicit and follow strict handling protocols. In contrast, LLMs exhibit implicit and unpredictable fault tolerance. Instead of producing explicit error messages, they attempt to generate a response, which may be misleading, subtly incorrect, or entirely hallucinated. This absence of strict error signaling complicates promptware debugging and reliability assessment. 

\noindent \textbf{C8. Execution Control:} Traditional runtime environments provide precise execution control, allowing engineers to step through code line by line, inspect variable states, and utilize debugging tools. In contrast, LLM execution is opaque—engineers can only influence behavior indirectly through prompt modifications. Debugging is heuristic and experimental, often relying on iterative prompt adjustments rather than structured debugging tools. This lack of systematic execution control makes diagnosing and fixing undesired behaviors more challenging.

\noindent \textbf{C9. Access Control:} Traditional runtime environments enforce strict access controls through permissions, memory protections, and sandboxing. In contrast, LLMs lack fine-grained access control mechanisms, making them vulnerable to security risks such as prompt injection and unintended leakage of sensitive information. Ensuring security in promptware is challenging due to LLMs’ inability to enforce rigid execution boundaries or restrict unauthorized data access.

\noindent \textbf{C10. Memory Management:} In traditional runtime environments, memory management is explicit, with direct control over allocation, deallocation, and garbage collection, ensuring efficient resource handling. In contrast, LLMs lack persistent memory in the conventional sense and process inputs contextually, meaning they do not retain state across interactions unless external memory mechanisms (e.g., databases) are used. For promptware, this creates a challenge in maintaining continuity and coherence across prompts, especially for tasks requiring long-term dependency tracking or sustained context across multiple interactions.

\begin{figure*}[t]
    \centering
    \resizebox{1\textwidth}{!}{
        \begin{forest}
            forked edges,
            for tree={
                grow=east,
                reversed=true,
                anchor=base west,
                parent anchor=east,
                child anchor=west,
                font=\large,
                rectangle,
                draw=black,
                rounded corners,
                base=left, 
                align=left,
                minimum width=4em,
                edge+={darkgray,line width=1pt},
                s sep=3pt,
                inner xsep=2pt,
                inner ysep=3pt,
                line width=1pt,
                ver/.style={rotate=90, child anchor=north, parent anchor=south, anchor=center},
            },
            where level=1{text width=10em, font=\normalsize}{},
            where level=2{text width=24em, font=\normalsize}{},
            where level=3{text width=11em, font=\normalsize}{},
            [Promptware Engineering
                [{\;}Prompt Requirements\\Engineering
                    [{\;}O1: LLM-driven prompt RE
                        [{\;}C4/C5/C6/C9]
                    ]
                    [{\;}O2: Functional and non-functional prompt requirements          [{\;}C1/C2/C3/C5/C7/C9]
                    ]
                    [{\;}O3: Multi-objective prompt requirements trade-off
                        [{\;}C4/C5/C8/C9]
                    ]
                    [{\;}O4: Ambiguity-resilient prompt specifications
                        [{\;}C2]
                    ]
                ]
                [{\;}Prompt Design
                    [{\;}O5: Prompt design patterns
                        [{\;}C1/C2/C3/C6/C7/C10]
                    ]
                    [{\;}O6: Prompt design tools
                        [{\;}C1]
                    ]
                    [{\;}O7: Prompt design metrics
                        [{\;}C1/C4/C5/C7/C10]
                    ]
                    [{\;}O8: Prompt pattern repositories
                        [{\;}C3/C6/C9/C10]
                    ]
                ]
                [{\;}Prompt Implementation
                    [{\;}O9: Prompt-centric programming languages
                        [{\;}C1/C2/C7]
                    ]
                    [{\;}O10: Prompt compilation
                        [{\;}C1/C2/C9]
                    ]
                    [{\;}O11: Prompt-centric IDEs
                        [{\;}C2/C3/C4/C7/C10]
                    ]
                    [{\;}O12: Prompt optimization
                        [{\;}C1/C2/C3]
                    ]
                    [{\;}O13: Online prompt implementation
                        [{\;}C2/C6/C10]
                    ]
                    [{\;}O14: Role-playing in prompts
                        [{\;}C5/C6]
                    ]
                    [{\;}O15: Prompt libraries and APIs
                        [{\;}C1/C2/C9]
                    ]
                ]
                [{\;}Prompt Testing and\\ Debugging 
                    [{\;}O16: Flaky test of prompts
                        [{\;}C2/C4/C5]
                    ]
                    [{\;}O17: Test input generation in prompt testing
                        [{\;}C1/C2/C4]
                    ]
                    [{\;}O18: Test oracle in prompt testing
                        [{\;}C1/C2/C3/C4]
                    ]
                    [{\;}O19: Test adequacy in prompt testing
                        [{\;}C1]
                    ]
                    [{\;}O20: Unit testing and integration testing of prompts
                        [{\;}C4/C8]
                    ]
                    [{\;}O21: Non-functional testing of prompts
                        [{\;}C5/C9]
                    ]
                    [{\;}O22: Prompt debugging
                        [{\;}C4/C6/C7/C8/C10]
                    ]
                ]
                [{\;}Prompt Evolution 
                    [{\;}O23: Evolution driven by code\texttt{,} LLM\texttt{,} and user feedback
                        [{\;}C1/C4/C5/C6]
                    ]
                    [{\;}O24: Versioning and traceability
                        [{\;}C3/C4/C6/C8/C9]
                    ]
                ]
                [{\;}Prompt Deployment \\and Monitoring 
                    [{\;}O25:  Deployment pipelines and automation
                        [{\;}C1/C2/C4/C6/C8]
                    ]
                    [{\;}O26: Runtime monitoring and observability
                        [{\;}C3/C4/C5/C6/C7/C9]
                    ]
                    [{\;}O27: Adaptive monitoring with feedback loops
                        [{\;}C2/C4/C5/C10]
                    ]
                ]
            ]
        \end{forest}
    }
    \caption{Roadmap for promptware engineering, highlighting key activities alongside associated research opportunities (O1, O2, etc.) and the relevant promptware characteristics (C1 to C10) to be considered for each opportunity.}
    \label{fig:roadmap}
\end{figure*}
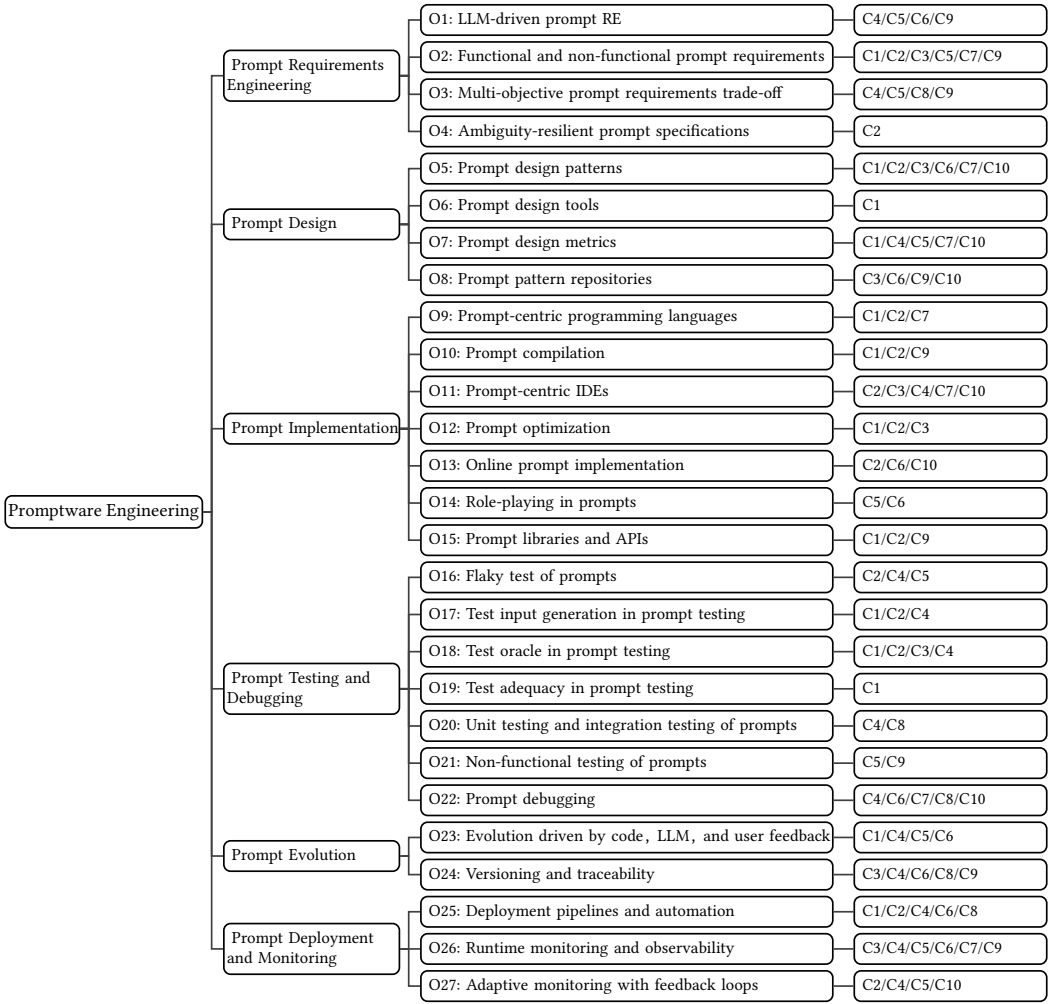

\section{Promptware Engineering: Roadmap}
This section presents a roadmap for promptware engineering, as shown in Figure~\ref{fig:roadmap}, covering key SE activities.

\subsection{Prompt Requirements Engineering}
Requirements engineering (RE) serves as the foundation of software development, focusing on translating user needs into clear, actionable specifications. In the realm of promptware engineering, RE involves identifying the requirements and defining the specifications for prompts.

\noindent \textbf{O1: LLM-driven prompt RE (C4, C5, C6, C9).} 
In traditional RE, requirements are primarily derived from user needs. However, in promptware engineering, requirements must also account for LLM capabilities and constraints, which influence how prompts are designed, refined, and validated.

First, prompt requirements must align with LLMs' reasoning ability, language comprehension, and domain expertise. However, LLMs' capability boundaries (C6) are unpredictable and evolve, necessitating adaptive requirements that can be iteratively refined. Additionally, LLMs exhibit non-deterministic execution (C4), meaning the same prompt can produce different outputs, complicating the definition of consistent and testable requirements.

Second, LLMs exhibit human-like characteristics (C5), such as personality~\cite{yaoqipersonality}, biases~\cite{WanWHGBL23}, emotional alignment~\cite{huang2024apathetic}, and contextual reasoning~\cite{giallanza2024context}. These characteristics must be explicitly accounted for in prompt requirements, especially in culturally sensitive or ethical applications. For example, requirements should define tone, formality, and ethical safeguards to ensure appropriate and unbiased interactions.

Furthermore, prompt requirements must also consider security risks of LLMs (C9). Poorly designed prompts may inadvertently expose sensitive information or be vulnerable to adversarial manipulation. 

To define effective prompt requirements, RE in promptware engineering can integrate insights from both computer and social sciences, fostering interdisciplinary collaboration among computer scientists, linguists, and psychologists.

\noindent \textbf{O2: Functional and non-functional prompt requirements (C1, C2, C3, C5, C7, C9).}
In traditional SE, functional requirements define the explicit tasks a system must perform, while non-functional requirements describe quality attributes such as performance, security, and robustness. This distinction remains essential in promptware engineering but requires adaptation due to the open-ended and ambiguous nature of natural language prompts (C1, C2).

Functional prompt requirements need to ensure clarity, specificity, and context-awareness. Prompts should be designed to explicitly communicate task objectives while minimizing ambiguities (C1, C2). Although deterministic correctness  for prompts is infeasible, ensuring reliability remains a core objective (C3).

Non-functional requirements, in contrast, ensure that prompts produce secure, fair, efficient, and robust behaviors. Prompts should be resistant to bias amplification, adversarial manipulation, and ethical risks (C5, C9). Additionally, performance constraints, such as response latency and token efficiency, must be balanced against quality and interpretability. As LLMs do not follow traditional error-handling mechanisms (C7), prompt robustness must be evaluated under adversarial conditions to prevent failures that might otherwise go unnoticed.

Due to the probabilistic and evolving nature of LLMs, the definition, scope, and prioritization of functional and non-functional requirements must remain adaptable and be continuously refined through empirical validation.

\noindent \textbf{O3: Multi-objective prompt requirements trade-off (C4, C5, C8, C9).} 
Promptware engineering often requires balancing competing objectives, making trade-off analysis a critical aspect of RE. For instance, role-playing techniques, which guide LLMs to assume specific roles, can enhance response relevance but may also amplify pre-existing biases~\cite{xinyuerole} (C5). Similarly, improving prompt clarity and specificity by providing detailed instructions enhances output quality but increases token consumption, impacting computational efficiency and cost.

Addressing these trade-offs necessitates a structured framework for systematically evaluating competing priorities. Since LLM execution is inherently non-deterministic (C4) and lacks direct execution control (C8), trade-off decisions need to rely on empirical testing and iterative refinement rather than formal verification. Additionally, prompt security (C9) must be integrated into trade-off considerations to prevent usability and performance optimizations from introducing vulnerabilities such as prompt injection attacks.

Developing systematic methodologies for balancing multiple prompt objectives remains a key research challenge, requiring the integration of multi-objective optimization techniques into promptware engineering workflows.

\noindent \textbf{O4: Ambiguity-resilient prompt specifications (C2).}
Natural language prompts are inherently ambiguous (C2), yet precise specification is crucial for achieving predictable and reliable performance. Traditional software specification techniques are inadequate for promptware engineering because they rely on formal syntax and semantics, whereas prompts depend on natural language, which lacks the rigor necessary for deterministic validation. This challenge highlights the need for new specification approaches that mitigate ambiguity while preserving the adaptability required for effective prompt design.

One potential direction is the development of semi-formal or formal specification languages specifically tailored for prompts. These languages could integrate structured templates with natural language annotations to improve clarity and reduce ambiguity. For example, structured elements could define task objectives, constraints, and contextual dependencies, while natural language annotations provide flexibility and interpretability. A hybrid approach like this could enable automated validation and consistency checks, ensuring that prompts adhere to defined requirements and perform reliably across varying conditions and use cases.

\subsection{Prompt Design}
Software design defines the architecture, components, interfaces, and characteristics of a system to ensure it meets specified requirements. In the context of promptware engineering, design focuses on structuring and organizing prompts to achieve effective model interactions.

The current prompt design has primarily been driven by the AI community. Common prompt structures, such as zero-shot (direct queries without examples)~\cite{wang2019survey}, few-shot (providing in-context examples)~\cite{BrownMRSKDNSSAA20}, Chain-of-Thought (CoT, breaking down reasoning into intermediate steps)~\cite{Wei0SBIXCLZ22}, Tree-of-Thought (explicitly exploring branching reasoning paths)~\cite{YaoYZS00N23}, self-consistency (sampling multiple reasoning paths and aggregating their outputs)~\cite{WSLCNCZ23}, and Retrieval-Augmented Generation (RAG) (incorporating external knowledge retrieval)~\cite{LewisPPPKGKLYR020}, are widely used but developed empirically rather than systematically.

SE has a long history of formalizing design patterns, i.e., reusable solutions to common design problems. A design pattern provides a high-level template for addressing specific challenges rather than a rigid implementation structure. This concept can be extended to promptware engineering, where recurring prompt structures and strategies can be classified as prompt design patterns, offering systematic frameworks for designing effective prompts.

\noindent \textbf{O5: Prompt design patterns (C1, C2, C3, C6, C7, C10).} 
The SE community can collaborate with AI researchers to move beyond ad hoc prompt templates and establish formalized prompt design patterns with well-defined specification structures. Rather than relying on informal analogies to classical SE patterns, such formalization would explicitly define a prompt pattern in terms of its intent, applicability conditions, constituent components (e.g., roles, instructions, examples, and constraints), and expected effects on model behavior.

A formal prompt pattern specification could further include constraints and invariants that the prompt structure is designed to enforce, as well as known failure modes and trade-offs. This aligns with emerging work on agentic design patterns~\cite{Antonio}, which demonstrates how interaction patterns for intelligent systems can be systematically described, analyzed, and reused. In particular, a formal prompt pattern may specify the interaction structure and information flow between prompt instances, such as how intermediate outputs are constrained, validated, and passed across steps, as exemplified by prompt chaining patterns in agentic systems. Adopting a similar specification discipline would allow prompt patterns to be reasoned about, compared, and incrementally refined, rather than treated as purely descriptive heuristics.

Formalizing prompt design patterns in this manner promotes standardization (C1), enabling consistent and reproducible prompt practices, and supports systematic optimization to improve prompt quality and reduce ambiguity (C2, C3). It can also help delineate LLM capability boundaries by clarifying which classes of tasks particular prompt structures are suitable for, and where they are likely to fail (C6).

Beyond codifying existing practices, there is an opportunity to discover new prompt design patterns through systematic experimentation, empirical evaluation, and longitudinal analysis of prompt usage. For example, recursive prompting, where the output of one interaction is explicitly constrained and fed into subsequent iterations, may constitute a distinct pattern for tasks requiring iterative refinement or sustained state tracking (C10). Similarly, failure-resilient prompt patterns that incorporate self-checking or fallback strategies could be formalized to mitigate hallucinations and improve robustness (C7).

\noindent \textbf{O6: Prompt design tools (C1).} 
Just as SE provides tools such as Unified Modeling Language (UML) diagrams to support the implementation of design patterns, the SE community can develop specialized tools for prompt design. These tools could assist promptware engineers in visualizing, assembling, and validating prompt structures (C1), streamlining the development process, and enhancing prompt effectiveness.

\noindent \textbf{O7: Prompt design metrics (C1, C4, C5, C7, C10).} 
In traditional SE, design quality is typically assessed using metrics such as cohesion, coupling, and complexity. Applying similar metrics to prompt design could establish a structured and rigorous evaluation framework for prompts. Cohesion could measure how effectively the components of a prompt work together to achieve the desired outcome. Coupling might evaluate the extent to which a prompt depends on external systems, such as retrieval-based components in RAG. Complexity could assess the structural intricacy of a prompt (C1) and the cognitive load it imposes on the LLM (C5).

Additional prompt-specific metrics could further enhance evaluation. A memory evaluation metric could quantify how well a prompt maintains context across multiple turns, ensuring critical information persists throughout an interaction (C10). A probabilistic determinism score could measure output variance for the same input, assessing reliability across different LLM versions and temperature settings (C4). Moreover, error-handling metrics could evaluate how effectively a prompt minimizes hallucinations and ambiguous responses (C7).

\noindent \textbf{O8: Prompt pattern repositories (C3, C6, C9, C10).} 
A shared repository of prompt design patterns, akin to the design pattern catalog in SE, could enhance innovation and standardization in promptware engineering. Such a repository would provide promptware engineers with a collection of well-documented, reusable solutions for common challenges, facilitating faster adaptation and iteration. By reducing reliance on trial-and-error approaches, this resource would promote consistency and improve prompt quality (C3).

Additionally, the repository could include capability-aware prompt patterns, documenting which designs are best suited for specific LLM functionalities (C6).

Security should also be a core component of this repository (C9). A dedicated section on adversarial robustness patterns could catalog techniques for mitigating prompt injection and unintended information leakage. 

Finally, patterns for long-term dependency tracking (e.g., summarization prompts that maintain context across multiple exchanges) could help address LLM statelessness, a key limitation in multi-turn interactions (C10).

\subsection{Prompt Implementation}\label{pimplem}
In traditional SE, the implementation phase focuses on writing programs that translate specifications into code. In the context of promptware engineering, implementation involves crafting prompts as natural language instructions to be processed by LLMs. Drawing from established SE practices, we highlight several research directions that could enhance the implementation of prompts. 

\noindent \textbf{O9: Prompt-centric programming languages (C1, C2, C7).} A key opportunity in promptware engineering lies in the design of prompt-centric programming languages that treat prompts as first-class program entities rather than unstructured natural-language strings. Such languages would provide dedicated primitives for expressing prompt templates, contextual dependencies (e.g., roles, constraints, and examples), and structured output contracts, enabling prompts to be composed, parameterized, and reused in a principled manner.
 
Beyond existing frameworks such as LangChain~\cite{oshin2025learning} or Liquid~\cite{liquiddoc} prompt templates, which focus on prompt orchestration and textual templating rather than language-level abstractions, prompt-centric languages could introduce formal programming abstractions, including modular prompt units, explicit context scoping, and typed or constraint-based output specifications, to support static reasoning and systematic error handling (C1, C2, C7). Unlike traditional programming languages, which assume deterministic semantics, these languages need to explicitly account for LLM non-determinism, for example, by adopting soft or refinement-based types, confidence-aware contracts, or property-level specifications that constrain acceptable outputs rather than prescribing exact values (C2). For instance, a prompt unit may define a task description and role context together with an output contract that specifies required structural or stylistic properties of the response, while allowing variability in the exact generated content.

From a language-design perspective, prompt-centric programming languages naturally favor declarative constructs~\cite{KhattabSMZSVHSJ24,XingLCHZSL25} for expressing user intent and output properties, while selectively incorporating imperative control for prompt orchestration and interaction flow, resulting in a hybrid paradigm suitable for promptware systems. Recent work on controlled natural languages for prompts, such as CNL-P~\cite{XingLCHZSL25}, demonstrates the effectiveness of declarative, grammar- and semantics-constrained prompts with static analysis support. Prompt-centric programming languages can build on and extend this line of work by situating such declarative prompt specifications within a broader SE context that supports modularity, composition, analysis, and long-term maintainability of promptware artifacts.

\noindent \textbf{O10: Prompt compilation (C1, C2, C9).}
Natural language is ambiguous and context-dependent (C2), making it difficult for LLMs to interpret prompts with precision. This challenge resembles a well-known problem in SE, where high-level programming languages must be transformed into structured, optimized machine code to ensure reliable execution. In traditional computing, compilers address this issue by systematically analyzing, refining, and restructuring code before it reaches the processor. Inspired by this concept, we envision prompt compilation, a process that translates human-written prompts into well-defined, optimized representations that LLMs can process more effectively.

Prompt compilation can involve multiple stages to ensure that a prompt is both information-dense and logically structured. First, the prompt undergoes lexical and syntactic analysis, where it is tokenized and parsed to extract key concepts, tasks, and dependencies. The next step involves creating a prompt intermediate representation, a structured, language-agnostic format that serves as an intermediary between raw prompts and LLM execution. This representation enables systematic analysis and transformation of the prompt, ensuring that it is logically sound and free from ambiguity (C2). After the creation of prompt intermediate representation, semantic optimization refines the structure further (C1), making implicit logic explicit, reorganizing sub-tasks, and converting natural language instructions into a clear execution plan. To improve efficiency, token compression eliminates redundancies, reformats expressions for clarity, and, when necessary, translates the prompt into a more compact form to reduce computational overhead. Finally, as part of the compilation process, security enhancements introduce constraints and safeguards (C9). These prevent prompt injection attacks and unintended outputs, ensuring that the prompt is not only efficient and logically sound but also secure during execution, much like how compilers ensure that the machine code is robust and free from errors before it runs on the processor.

\noindent \textbf{O11: Prompt-centric IDEs (C2, C3, C4, C7, C10).} In traditional SE, Integrated Development Environments (IDEs) facilitate coding, debugging, and testing. A similar opportunity exists in promptware engineering to develop prompt-centric IDEs, which is also discussed in a recent paper~\cite{hassan2024rethinking}. These specialized tools could offer features like real-time feedback on prompt performance, automatic generation of prompt variations, and integration with prompt evaluation metrics (C3). By streamlining prompt creation and refinement, such tools could significantly reduce the time and effort required to implement high-quality prompts.

Given the non-deterministic nature of LLM outputs (C4), the IDE could incorporate features that assess probabilistic consistency across prompt outputs, helping engineers track variability and ensure reproducibility. The IDE could also integrate contextual checks to identify ambiguities in prompts (C2), offering error messages to inform potential issues (C7) and suggestions to make them more explicit and deterministic. Additionally, integration with external memory mechanisms (C10) would help maintain the coherence of prompts over time, especially in long-term interactions

\noindent \textbf{O12: Prompt optimization (C1, C2, C3).}
Prompt optimization is a significant challenge in promptware engineering. Unlike traditional software code, which benefits from established analysis and optimization tools, prompts are highly context-sensitive (C2), requiring a more nuanced approach to optimization. The outputs of prompts can be influenced by subtle variations in wording (C3), context, and structure, making it difficult to ensure consistency and effectiveness across a broad range of tasks. In prompt optimization, factors such as structure, tone, contextual information, and task-specific instructions must be fine-tuned to improve outputs (C1, C2).

To address these challenges, search-based SE methodologies~\cite{harman2001search} offer promising solutions. Search-based SE techniques, such as genetic algorithms, can automatically explore and evaluate different prompt configurations to identify the optimal combination of elements (such as wording, structure, and context) that maximizes effectiveness across various tasks. By iterating through diverse variations, these methods enhance robustness and mitigate the impact of subtle prompt differences. Additionally, context-aware optimization algorithms can dynamically adjust the prompt elements in real-time, ensuring that the prompt remains effective even as context or task requirements change. Furthermore, multi-objective optimization frameworks can help balance competing factors such as accuracy, consistency, and efficiency, providing a comprehensive solution for prompt optimization.

\noindent \textbf{O13: Online prompt implementation (C2, C6, C10).} Prompt development frequently requires online implementation, where responses are generated in real time with minimal delay to ensure prompt user feedback~\cite{AlshahwanHHMSW24}. The need for online prompts stems from the dynamic runtime environment (C6), which contrasts with the more rigid, controlled setup of traditional offline systems. While online systems are preferred for their responsiveness, the transition from offline to online frameworks is not always feasible due to the continuous adaptability required in real-time contexts.

The primary challenge of online prompt implementation lies in effectively balancing real-time adaptability with contextual coherence and response accuracy. Unlike offline systems that rely on static prompts, online systems must continuously evolve in response to user input, system behavior, and fluctuating environmental factors. This inherent dynamism introduces difficulties in maintaining the relevance and consistency of prompts, particularly when handling large-scale or complex interactions. This area presents significant research opportunities, especially in developing algorithms capable of learning from real-time interactions. Key avenues of exploration include adaptive prompt generation, improving context-aware language understanding (C2), and advancing long-term memory mechanisms that allow models to retain and recall context across multiple sessions (C10).

\noindent \textbf{O14: Role-playing in prompts (C5, C6).}
Role-playing is a widely adopted prompting strategy for enhancing the utility of LLMs by simulating real-world roles~\cite{tseng2024two}.
A key challenge in role-playing prompts is specifying optimal roles, which requires a deep understanding of LLMs' human-like characteristics and their capabilities in role-playing (C5). This understanding is crucial for ensuring better contextual understanding and engagement. Defining these roles also demands a balance between clarity and flexibility—too rigid a role can limit the model's responses, while too vague a role can lead to loss of context. Additionally, the role must be tailored to LLMs' capabilities (C6), ensuring that it can effectively handle complex and dynamic interactions.

Specifying social roles in prompts can also introduce biases linked to stereotypes present in training data~\cite{xinyuerole}. This raises a significant issue: how can we use role-playing to improve interaction while minimizing bias? Solutions could include refining role definitions, applying bias-mitigation techniques, and continuously evaluating prompts across diverse demographic and cultural contexts. Ethical guidelines and fairness-aware mechanisms should be integrated to ensure role-playing adds value without reinforcing harmful stereotypes.

\noindent \textbf{O15: Prompt libraries and APIs (C1, C2, C9).} 
In SE, reusable code libraries have significantly enhanced development efficiency. A similar approach in promptware engineering could involve curated libraries of reusable prompt templates for common tasks such as summarization, question answering, and translation. These libraries provide baseline prompts that engineers could easily adapt to suit specific needs. To ensure standardization, these libraries would need to follow consistent structures (C1) and allow for easy customization while minimizing ambiguity (C2).

However, similar to the open-source software ecosystem in SE, prompt engineering must address challenges related to copyright and licensing. Recent research~\cite{abs240515161,abs240219200} has highlighted the need for frameworks that can define, detect, and resolve intellectual property issues. Developing licensing models that strike a balance between open access, commercial use, and ethical considerations will be crucial for the sustainable growth of promptware engineering.

Standardized prompt APIs for interacting with prompts and LLMs could further streamline integration into larger software ecosystems. These APIs could include methods for prompt assembly and output validation, enabling modular, maintainable, and scalable prompt-based applications. To ensure the security of prompt execution and protect sensitive data, these APIs should incorporate access control mechanisms (C9) that define who can modify or access specific prompt configurations. Additionally, these standardized APIs could support dependency tracking across reusable prompts, improving both the modularity and traceability of the promptware engineering process.

\subsection{Prompt Testing and Debugging}
Software testing verifies whether software behaves as expected, while debugging identifies and resolves issues. In promptware engineering, prompt testing and debugging aim to detect and correct undesired behaviors triggered by prompts. 
Unlike traditional testing and debugging, which focus on deterministic code, prompt testing and debugging need to handle non-deterministic outputs and ambiguous specifications, introducing unique challenges.

\noindent \textbf{O16: Flaky test of prompts (C2, C4, C5).}
In SE, a flaky test refers to a test that produces inconsistent results, yielding failures or successes unpredictably across multiple runs without changes to the underlying code~\cite{luo2014empirical}. This issue is particularly prevalent in prompt testing, where the probabilistic nature of natural language (C2), combined with the non-deterministic outputs of LLMs (C4), creates challenges in ensuring consistent test results.

To determine whether a test passes or fails, multiple attempts may be necessary, relying on aggregated or consensus-based assessments to distinguish between random fluctuations and genuine prompt flaws. One potential solution is to move beyond the traditional binary pass/fail approach, introducing a success threshold (e.g., requiring outputs to be highly similar in at least 80\% of test runs). While this reduces the impact of random variations, it does not entirely eliminate the uncertainty associated with flaky tests.

To mitigate randomness, researchers often adjust the temperature parameter, which influences the level of creativity and variability in the generated text~\cite{ouyang2024empirical}. A temperature of zero minimizes randomness, yielding more deterministic outputs. However, even with this setting, LLMs still exhibit some degree of variability~\cite{ouyang2024empirical}. Moreover, real-world applications typically require higher temperature settings to encourage more dynamic and creative responses (C5). Testing with a temperature of zero may not fully represent real-world scenarios, leaving flaky test in prompt testing as a significant challenge. Flaky test for prompts remains an open problem, and further research is needed to develop more dependable solutions.

\noindent \textbf{O17: Test input generation in prompt testing  (C1, C2, C4).} 
Test input for prompt testing refers to the specific values assigned to different elements (i.e., variables) within a prompt. In practice, prompts may receive inputs from both LLMs and other software components. For example, in a code generation scenario, a prompt might include a task description provided by other components, as well as feedback or contextual information from prior LLM-generated outputs. This interaction introduces challenges for test input generation in prompt testing.

Specifically, it becomes difficult to generate representative and comprehensive test data that covers the range of possible inputs the prompt might encounter during actual usage (C1). The dynamic nature of inputs, such as changing task descriptions or evolving context from prior LLM outputs (C2), means that test inputs must account for a wide variety of conditions. Additionally, the interdependence between prompts and external software components adds another layer of complexity, as testing requires ensuring that both the prompt and the surrounding system behave cohesively and predictably.

Further complicating matters, the variability introduced by previous LLM outputs (which may differ slightly each time, due to factors like randomness) makes it hard to ensure consistent test coverage (C4). Test inputs need to be designed to capture this stochastic behavior while still allowing for reliable evaluation of the prompt’s functionality. This dynamic and multifaceted nature of prompt testing demands robust input generation strategies that can handle the unpredictability of real-world interactions.

\noindent \textbf{O18: Test oracle in prompt testing (C1, C2, C3, C4).}
The test oracle problem, which involves determining the correct behavior in response to an input, is a well-known challenge in software testing~\cite{tseBarrHMSY15}. In prompt testing, a test oracle is a mechanism or reference that assesses whether the output triggered by a prompt with a given test input is correct or acceptable. Unlike traditional testing, where the expected output is typically clear and deterministic, prompt testing faces a greater challenge due to the inherent subjectivity and open-ended nature of many LLM prompts (C1, C2). For example, prompts might ask the model to provide helpful or polite responses or elicit creative problem-solving. These specifications are often vague and context-dependent, complicating the determination of a `correct' output (C3).

In current promptware practices, manual evaluation is often used to assess output correctness~\cite{nahar2024beyond}. While effective, this method is time-consuming and susceptible to human biases. Techniques like multi-reviewer consensus or double-blind evaluations can mitigate these biases by offering more balanced judgments. An alternative, the LLM-as-a-judge approach, uses one LLM to evaluate another’s output~\cite{chenetal2024humans}. While this method can expedite testing by filtering out clearly incorrect or inappropriate responses, it may amplify biases if the models share similar training data or limitations. Therefore, human oversight remains essential, particularly in high-stakes contexts.

Metamorphic testing is a common technique used to address the traditional test oracle problem. This method defines relationships, known as metamorphic relations, between variations in input and expected output consistency~\cite{csurChenKLPTTZ18}. For example, when testing a prompt for code generation, rephrasing the task description should result in consistent model responses if the prompt is well-constructed. If significant discrepancies arise between the rephrased inputs and outputs, it may indicate a flaw in the prompt. However, because LLMs often produce different outputs for the same prompt (C4), it is difficult to assess whether the variation is due to prompt inconsistencies or model behavior. To address this, research can focus on developing metamorphic testing techniques specifically designed to test how small, controlled changes in input influence outputs, ensuring that expected behavior remains consistent across variations.

\noindent \textbf{O19: Test adequacy in prompt testing (C1).}
Test adequacy is a key concept in traditional software testing, used to assess the coverage provided by existing tests~\cite{gligoric2013comparing}. In this context, it measures how thoroughly various aspects of the software, such as different code paths or functional areas, are tested. In prompt testing, adequacy shifts focus to evaluating how well different components of a prompt, such as task descriptions, context, examples, and formatting requirements, are covered. Unlike traditional testing, which often relies on metrics like branch coverage, these methods are not directly applicable to prompt testing due to the unstructured and flexible nature of natural language. One way to improve test adequacy is by developing formalized structures (C1) for prompt components, which would provide a clearer basis for establishing coverage metrics.

\noindent \textbf{O20: Unit testing and integration testing of prompts (C4, C8).} 
Unit testing involves evaluating individual prompts or closely related sets of prompts in isolation to ensure that each prompt functions as intended. This approach verifies that each prompt meets its specific requirements without interference from other prompts or external contexts. However, in complex LLM-based software, prompts are often chained together~\cite{PromptsArePrograms}, as in conversation flows where each response depends on previous inputs. Integration testing ensures that these interconnected prompts produce coherent, accurate, and contextually consistent outputs when used together. Errors that may go undetected during unit testing can surface only when prompts interact in real-world conditions, highlighting the importance of integration testing to verify the system's overall behavior. Since determinism and execution control are often difficult to guarantee in LLMs (C4, C8), enhancing integration testing might involve developing tools that give developers more control over the flow of execution, including debugging tools that make internal processes more observable. This will help identify where unexpected results emerge during prompt interactions.

\noindent \textbf{O21: Non-functional testing of prompts (C5, C9):} As prompt-enabled systems are increasingly adopted in human-centric applications, ensuring that non-functional requirements such as fairness, security, and privacy are met is becoming critically important. Non-functional testing helps verify that these aspects are addressed thoroughly.

\emph{Fairness testing.} Since LLMs exhibit human-like characteristics (C5), they may also display human-like social biases or discriminatory tendencies when prompted with inappropriate inputs~\cite{xinyuerole,WanWHGBL23}. These biases can lead to the unfair treatment of specific groups or perpetuate harmful stereotypes related to various roles. If left unchecked, the repeated use of biased responses can normalize these stereotypes, subtly influencing public perception and reinforcing social inequalities. Fairness testing aims to identify biases caused by prompts. 

\emph{Security testing.} Security testing aims to identify vulnerabilities that could compromise the safety and integrity of prompt-enabled systems. For example, adversarial attacks like prompt injection (C9) exploit weaknesses by crafting malicious inputs to manipulate the LLM or bypass safeguards~\cite{debenedetti2024agentdojo}. 

\emph{Privacy testing.} Given that prompt-enabled systems often handle sensitive information, ensuring privacy is crucial. Prompts must undergo rigorous testing to ensure they do not inadvertently expose private or confidential data (C9). For example, if a prompt prompts an LLM to disclose sensitive information, whether from user input or internal training data, it constitutes a serious privacy breach. 

\noindent \textbf{O22: Prompt debugging (C4, C6, C7, C8, C10).}
Poorly constructed or ambiguous prompts may result in incorrect or undesirable outputs. Prompt debugging focuses on identifying and resolving these issues.

A primary challenge in prompt debugging is reproducing bugs, as LLMs are often black-box and stateless, exhibiting flakiness, i.e., identical prompts can produce inconsistent outputs (C4). This variability complicates bug reproduction, especially when context, recent interactions, or model updates change outputs. To address this, robust debugging mechanisms are essential. Tools that capture and replay the system’s state during prompt execution can help developers reproduce bugs and analyze discrepancies across different configurations. In this context, state capture does not aim to access internal states of the LLM, which can be inherently opaque, but rather to record an externally observable prompt execution state. This state includes the prompt artifact (e.g., template, instructions, and exemplars), runtime configurations (e.g., model version and decoding parameters), and interaction context (e.g., conversation history and intermediate tool outputs). Capturing these elements defines a versioned, replayable execution boundary, enabling controlled re-execution despite the black-box and stateless nature of LLMs.

Once a bug is reproduced, identifying its root cause, i.e., whether in the prompt design or the model’s behavior, becomes crucial. The black-box nature of LLMs (C6) complicates this, as developers lack visibility into the model’s internal processes. Additionally, the absence of execution control (C8) means developers cannot inspect intermediate states. As a result, prompt debugging often relies on indirect techniques like prompt decomposition and ablation studies, modifying parts of the prompt to isolate the cause. 
However, in multi-turn or multi-step prompts, the inherent flakiness of LLMs (C4) further complicates localization across runs, as the same prompt may yield different outputs under identical conditions.
To address this challenge, debugging should be conducted at the level of a prompt debugging session, defined as a bounded, versioned sequence of prompt executions in which developers iteratively modify prompt components while holding the captured execution state fixed. This session-based abstraction does not eliminate flakiness, but it provides a structured framework to track and compare multiple executions, enabling statistical reasoning to distinguish genuine faults from stochastic variation. By organizing the debugging process at the session level, failures can be systematically analyzed across interactions rather than treating each prompt invocation in isolation.

After localization, the next challenge is bug fixing, where the problematic section of the prompt is adjusted while preserving its intended functionality. While bug fixes may resolve one issue, they can inadvertently introduce new ones, especially in multi-step or CoT prompts. To mitigate this, automated tools should be implemented to suggest and verify prompt modifications based on known patterns of failure. This pattern-based approach would help developers apply proven solutions to recurring problems, thereby streamlining the debugging process and reducing errors.

Additionally, incorporating error-handling mechanisms within LLMs could enhance their ability to provide explicit feedback when an issue arises. Instead of leaving developers to infer the problem, LLMs could flag uncertainties or ambiguities in the prompt and suggest improvements. This would address the lack of structured error messages (C7) and make the identification and resolution of issues more efficient.

Finally, for complex, multi-step prompts, maintaining continuity across interactions is crucial. Since LLMs lack persistent memory (C10), external mechanisms for contextual memory could be used to maintain coherence across multiple exchanges. This would help reduce errors caused by the loss of context and ensure smoother debugging, particularly for longer chains of reasoning.

\subsection{Prompt Evolution}
Software evolution typically involves iterative development and updates. Similarly, prompts require proactive adaptation and continuous refinement, especially in dynamic environments.

\noindent \textbf{O23: Evolution driven by code, LLM, and user feedback (C1, C4, C5, C6).}
Prompt evolution in LLM-based software is influenced by code changes, LLM updates, and user feedback. Unlike programming languages, prompts lack explicit syntax rules, making adaptation more challenging (C1). Additionally, LLMs generate responses probabilistically, complicating consistent behavior from prompt modifications (C4). As LLMs evolve with updates and new training data, prompts must be revalidated to prevent performance degradation (C6). LLMs’ human-like reasoning (C5) also introduces susceptibility to implicit biases and shifting linguistic patterns, requiring tracking and adaptation. 

To streamline prompt evolution, researchers and practitioners can develop context-aware prompt management systems that automatically detect shifts in code, model behavior, and user feedback, adjusting prompts accordingly. 
One approach is to introduce structured prompt templates and metadata tracking to improve control over prompt modifications. Additionally, automated validation mechanisms can assess whether prompt adjustments yield reliable results.

\noindent  \textbf{O24: Versioning and traceability (C3, C4, C6, C8, C9).}
In traditional SE, version control systems like Git are essential for tracking code changes and maintaining traceability. Similarly, promptware engineering would benefit from specialized version control tools to track prompt iterations, document modifications, and ensure accountability (C3). Such systems would enable engineers to compare prompt versions, ensuring consistent refinement and structured evolution.

Since prompts interact dynamically with evolving LLMs (C6), maintaining compatibility between prompt versions and software upgrades is critical. Any modifications, especially those prompted by LLM updates, code changes, user feedback, or shifting use cases, should be thoroughly documented. Given the non-deterministic nature of LLM execution (C4), precise versioning strategies are necessary to mitigate unintended variations in model behavior.

To support these, prompt versioning systems should incorporate detailed changelogs, update metadata, and explicit compatibility requirements, mirroring the principles of semantic versioning in traditional software. Additionally, automated diff-checking mechanisms could highlight changes in prompts and their impact on LLM responses, improving reliability.

Furthermore, access control mechanisms should be integrated into prompt management systems (C9) to ensure that modifications are authorized and traceable, preventing unauthorized edits or security vulnerabilities. Since prompt debugging remains experimental and lacks structured debugging tools (C8), versioning should support rollback mechanisms to restore previously stable versions when regressions occur.

\subsection{Prompt Deployment and Monitoring}
Modern SE increasingly emphasizes the integration of development and operations (DevOps), ensuring that applications are not only well-designed but also reliably deployed, monitored, and continuously improved. Similarly, promptware engineering must address deployment and runtime challenges, where the interplay of LLM characteristics and prompt design introduces unique research opportunities~\cite{TantithamthavornPKC25,diaz2024large}.

\noindent \textbf{O25: Deployment pipelines and automation (C1, C2, C4, C6, C8).}  
The lack of formal syntax and structure in prompts (C1) and their inherent ambiguity (C2) complicate systematic pre-deployment testing and validation, making it difficult to ensure that prompt outputs meet intended goals. To mitigate these issues, CI/CD-like pipelines can be designed for promptware that integrate automated testing tailored to prompt semantics and contextual requirements. Non-deterministic LLM outputs (C4) and unclear capability boundaries (C6) further challenge consistent behavior across deployments, suggesting the need for controlled rollout strategies such as canary deployments, staged releases, and safe rollback mechanisms. Limited execution control (C8) motivates the development of structured prompt templates and metadata tracking, enabling reproducible deployment, traceable modifications, and robust management of prompt evolution in production.

\noindent \textbf{O26: Runtime monitoring and observability (C3, C4, C5, C6, C7, C9).}  
Because correctness and quality are difficult to define for prompts (C3), and LLM outputs are non-deterministic (C4) while exhibiting human-like reasoning behaviors (C5), traditional monitoring techniques are insufficient to capture anomalies, biases, or unexpected interpretations. Establishing thresholds for normal versus abnormal behavior is complicated by unclear capability boundaries (C6) and the absence of explicit error signaling (C7). 

At the same time, limited access control (C9) exposes promptware to concrete operational and security threats, including prompt injection attacks, unintended data leakage, and jailbreak behaviors. For example, prompt injection may cause an LLM to override system instructions; data leakage may occur when sensitive context is unintentionally exposed in model outputs; and jailbreaks can bypass safety or compliance constraints at runtime.

These threats motivate the need for runtime observability mechanisms that go beyond passive logging. In practice, prompt-enabled systems can integrate guardrails and policy checks to validate inputs and outputs, role-based access control to restrict prompt modifications, and audit logs to track prompt evolution and usage. Observability frameworks can continuously monitor outputs for functional, performance, reliability, and security metrics, and surface trustworthiness indicators such as stability, fairness, safety, and compliance via dashboards and automated alerts.

Sandboxing, automated red-teaming, and security-focused monitoring techniques can further complement these frameworks, enabling real-time detection and mitigation of operational and security issues.

\noindent \textbf{O27: Adaptive monitoring with feedback loops (C2, C4, C5, C10).}  
Ambiguity and context dependence in prompts (C2), combined with LLM non-determinism (C4) and human-like reasoning (C5), make promptware behavior dynamic and sensitive to runtime conditions. The absence of persistent memory (C10) further complicates long-term tracking and adaptation. To address these challenges, adaptive monitoring systems can incorporate user feedback, runtime performance data, and error signals to drive semi-automated prompt refinement. By leveraging feedback loops and external memory mechanisms, these systems can adjust prompt behavior dynamically, maintain stable performance over time, and support continuous improvement across deployments.

\section{Limitations and Future Work}
As a vision paper, this work primarily focuses on articulating the conceptual framework for promptware engineering and outlining associated research opportunities. Several limitations arise naturally from this scope.

First, the proposed framework is largely theoretical and has not been implemented or empirically validated. Its practical effectiveness, scalability, and usability remain to be demonstrated in real-world prompt-enabled systems. Conducting systematic case studies, experiments, or deployment trials is beyond the scope of the current paper but represents an important avenue for future work. Empirical validation could evaluate the effectiveness of prompt development practices, deployment pipelines, runtime monitoring, versioning strategies, and other lifecycle activities, providing concrete evidence of the benefits, limitations, and trade-offs of promptware engineering.

Although we do not empirically validate the framework, we provide an illustrative vignette that walks through the promptware engineering lifecycle. Consider a prompt-enabled customer support system that uses an LLM to answer refund and delivery-related user inquiries. In the requirements phase, developers specify functional goals (e.g., generating accurate and polite responses), non-functional constraints (e.g., response latency and per-query cost), and trustworthiness requirements (e.g., avoiding biased or unsafe content). During design, these requirements are translated into structured prompt templates that separate system instructions, task descriptions, and contextual inputs, and define interactions with retrieved knowledge or backend tools. In implementation, prompts are instantiated, parameterized, and versioned as first-class artifacts alongside code. Testing evaluates prompts using automatically generated test cases, including edge cases and adversarial inputs, together with test oracles that assess correctness, robustness, and fairness. After deployment, runtime monitoring continuously observes prompt behavior by tracking output quality, stability, and policy compliance, logging prompt changes, and triggering alerts upon anomalies. As requirements evolve or failures are observed, prompts are iteratively refined and redeployed, completing a full promptware engineering lifecycle.

Second, effective application of the framework requires expertise spanning both prompt engineering and traditional SE. Promptware engineers must understand not only how to craft effective natural language prompts but also how to integrate them into a full software engineering lifecycle. Achieving this integration demands a combination of linguistic, technical, and process-oriented skills, which may not be readily available in all teams or organizations. Small teams or organizations with limited resources may face challenges in adopting systematic promptware engineering practices, potentially limiting the framework’s applicability in certain contexts. Future research could explore methods to lower the technical barrier, including automated tooling, guided prompt development environments, and training frameworks for multidisciplinary teams.

By acknowledging these limitations and outlining potential directions for support mechanisms, we aim to provide a balanced perspective on promptware engineering while guiding the community toward bridging the gap between conceptual vision and practical realization.

\section{Conclusion}
In this paper, we highlight 10 unique characteristics of promptware, which arise from its natural language programming and the use of LLMs as runtime environments, in contrast to traditional software paradigms. These characteristics introduce specific challenges that current experimental, trial-and-error practices fail to adequately address. To overcome these limitations, we propose promptware engineering, a systematic methodology that integrates established SE principles into prompt creation and optimization, moving beyond the ad hoc methods currently in use. Without the SE discipline, prompt development is likely to remain mired in trial-and-error. To support our vision, we present a roadmap with 27 research opportunities across critical activities such as prompt requirements engineering, design, implementation, testing, debugging, evolution, deployment, and monitoring. We argue that promptware engineering is inherently interdisciplinary, sitting at the intersection of SE and AI, and requires close collaboration between the two communities.

\section*{Acknowledgments}
This research is supported by the National Natural Science Foundation of China (Grant No. 62325201); by the ITEA grants GreenCode (Project No. 23016) and GENIUS (Project No. 23026); by the National Research Foundation, Singapore, and DSO National Laboratories under the AI Singapore Programme (AISG Award No. AISG2-RP-2020-019); and by the National Research Foundation, Singapore, and the Cyber Security Agency of Singapore under the National Cybersecurity R\&D Programme (NCRP25-P04-TAICeN). This research is also part of the IN-CYPHER Programme and is supported by the National Research Foundation, Prime Minister’s Office, Singapore, under its Campus for Research Excellence and Technological Enterprise (CREATE) Programme. Any opinions, findings, conclusions, or recommendations expressed in this paper are those of the authors and do not reflect the views of the supporting agencies.

\bibliographystyle{ACM-Reference-Format}
\bibliography{vision}

@misc{liquiddoc,
  title = {Liquid prompt’s documentation},
  howpublished = {\url{https://liquidprompt.readthedocs.io/en/stable/}},
  year = {2025}
}

@article{tseBarrHMSY15,
  author       = {Earl T. Barr and
                  Mark Harman and
                  Phil McMinn and
                  Muzammil Shahbaz and
                  Shin Yoo},
  title        = {The oracle problem in software testing: {A} survey},
  journal      = {{IEEE} Transactions on Software Engineering},
  volume       = {41},
  number       = {5},
  pages        = {507--525},
  year         = {2015}
}

@inproceedings{abs240219200,
  author       = {Yong Yang and
                  Changjiang Li and
                  Qingming Li and
                  Oubo Ma and
                  Haoyu Wang and
                  Zonghui Wang and
                  Yandong Gao and
                  Wenzhi Chen and
                  Shouling Ji},
  title        = {{PRSA:} Prompt stealing attacks against real-world prompt services},
  booktitle    = {Proceedings of the 34th {USENIX} Security Symposium, {USENIX} Security 2025},
  pages        = {2283--2302},
  year         = {2025}
}

@article{abs240515161,
  author       = {Huali Ren and
                  Anli Yan and
                  Chong{-}zhi Gao and
                  Hongyang Yan and
                  Zhenxin Zhang and
                  Jin Li},
  title        = {Are you copying my prompt? Protecting the copyright of vision prompt
                  for VPaaS via watermark},
  journal      = {Computer Standards \& Interfaces},
  volume       = {94},
  pages        = {103992},
  year         = {2025}
}

@inproceedings{LewisPPPKGKLYR020,
  author       = {Patrick S. H. Lewis and
                  Ethan Perez and
                  Aleksandra Piktus and
                  Fabio Petroni and
                  Vladimir Karpukhin and
                  Naman Goyal and
                  Heinrich K{\"{u}}ttler and
                  Mike Lewis and
                  Wen{-}tau Yih and
                  Tim Rockt{\"{a}}schel and
                  Sebastian Riedel and
                  Douwe Kiela},
  title        = {Retrieval-augmented generation for knowledge-intensive {NLP} tasks},
  booktitle    = {Proceedings of Advances in Neural Information Processing Systems 33: Annual Conference
                  on Neural Information Processing Systems 2020, NeurIPS 2020},
  year         = {2020}
}

@inproceedings{Wei0SBIXCLZ22,
  author       = {Jason Wei and
                  Xuezhi Wang and
                  Dale Schuurmans and
                  Maarten Bosma and
                  Brian Ichter and
                  Fei Xia and
                  Ed H. Chi and
                  Quoc V. Le and
                  Denny Zhou},
  title        = {Chain-of-thought prompting elicits reasoning in large language models},
  booktitle    = {Proceedings of Advances in Neural Information Processing Systems 35: Annual Conference
                  on Neural Information Processing Systems 2022, NeurIPS 2022},
  year         = {2022}
}

@inproceedings{BrownMRSKDNSSAA20,
  author       = {Tom B. Brown and
                  Benjamin Mann and
                  Nick Ryder and
                  Melanie Subbiah and
                  Jared Kaplan and
                  Prafulla Dhariwal and
                  Arvind Neelakantan and
                  Pranav Shyam and
                  Girish Sastry and
                  Amanda Askell and
                  Sandhini Agarwal and
                  Ariel Herbert{-}Voss and
                  Gretchen Krueger and
                  Tom Henighan and
                  Rewon Child and
                  Aditya Ramesh and
                  Daniel M. Ziegler and
                  Jeffrey Wu and
                  Clemens Winter and
                  Christopher Hesse and
                  Mark Chen and
                  Eric Sigler and
                  Mateusz Litwin and
                  Scott Gray and
                  Benjamin Chess and
                  Jack Clark and
                  Christopher Berner and
                  Sam McCandlish and
                  Alec Radford and
                  Ilya Sutskever and
                  Dario Amodei},
  title        = {Language models are few-shot learners},
  booktitle    = {Proceedings of Advances in Neural Information Processing Systems 33: Annual Conference
                  on Neural Information Processing Systems 2020, NeurIPS 2020},
  year         = {2020}
}

@article{wang2019survey,
  title={A survey of zero-shot learning: Settings, methods, and applications},
  author={Wang, Wei and Zheng, Vincent W and Yu, Han and Miao, Chunyan},
  journal={ACM Transactions on Intelligent Systems and Technology},
  volume={10},
  number={2},
  pages={1--37},
  year={2019}
}

@article{liu2024large,
  title={Large language model-based agents for software engineering: A survey},
  author={Liu, Junwei and Wang, Kaixin and Chen, Yixuan and Peng, Xin and Chen, Zhenpeng and Zhang, Lingming and Lou, Yiling},
  journal={ACM Transactions on Software Engineering and Methodology},
  pages = {accepted to appear},
  year={2024}
}

@article{hou2024large,
  title={Large language models for software engineering: A systematic literature review},
  author={Hou, Xinyi and Zhao, Yanjie and Liu, Yue and Yang, Zhou and Wang, Kailong and Li, Li and Luo, Xiapu and Lo, David and Grundy, John and Wang, Haoyu},
  journal={ACM Transactions on Software Engineering and Methodology},
  volume={33},
  number={8},
  pages={1--79},
  year={2024}
}

@article{wang2024software,
  title={Software testing with large language models: Survey, landscape, and vision},
  author={Wang, Junjie and Huang, Yuchao and Chen, Chunyang and Liu, Zhe and Wang, Song and Wang, Qing},
  journal={IEEE Transactions on Software Engineering},
volume={50},
  number={4},
  pages={911--936},
  year={2024}
}

@inproceedings{FanGHLSYZ23,
  author       = {Angela Fan and
                  Beliz Gokkaya and
                  Mark Harman and
                  Mitya Lyubarskiy and
                  Shubho Sengupta and
                  Shin Yoo and
                  Jie M. Zhang},
  title        = {Large language models for software engineering: Survey and open problems},
  booktitle    = {Proceedings of {IEEE/ACM} International Conference on Software Engineering: Future
                  of Software Engineering, ICSE-FoSE 2023},
  pages        = {31--53},
  year         = {2023}
}

@misc{metaPEdefine,
  title = {Prompting},
  howpublished = {\url{https://www.llama.com/docs/how-to-guides/prompting/}},
  year = {2025}
}

@misc{openaiPEdefine,
  title = {Prompt engineering best practices for ChatGPT},
  howpublished = {\url{https://help.openai.com/en/articles/10032626-prompt-engineering-best-practices-for-chatgpt}},
  year = {2025}
}

@misc{openaiguide,
  title = {Enhance results with prompt engineering strategies},
  howpublished = {\url{https://platform.openai.com/docs/guides/prompt-engineering}},
  year = {2025}
}

@inproceedings{parnin2023building,
  title={Building your own product copilot: Challenges, opportunities, and needs},
  author={Parnin, Chris and Soares, Gustavo and Pandita, Rahul and Gulwani, Sumit and Rich, Jessica and Henley, Austin Z},
  booktitle    = {Proceedings of the {IEEE} International Conference on Software Analysis, Evolution and
                  Reengineering, {SANER} 2025},
  pages        = {338--348},
  year         = {2025}
}

@article{bi2024deepseek,
  title={DeepSeek LLM: Scaling open-source language models with longtermism},
  author={Bi, Xiao and Chen, Deli and Chen, Guanting and Chen, Shanhuang and Dai, Damai and Deng, Chengqi and Ding, Honghui and Dong, Kai and Du, Qiushi and Fu, Zhe and others},
  journal={arXiv preprint arXiv:2401.02954},
  year={2024}
}

@inproceedings{dolata2024development,
  title={Development in times of hype: How freelancers explore generative AI?},
  author={Dolata, Mateusz and Lange, Norbert and Schwabe, Gerhard},
  booktitle={Proceedings of the IEEE/ACM 46th International Conference on Software Engineering, ICSE 2024},
  pages={1--13},
  year={2024}
}

@article{liu2025first,
  title={A First Look at Bugs in LLM Inference Engines},
  author={Liu, Mugeng and Zhong, Siqi and Bi, Weichen and Zhang, Yixuan and Chen, Zhiyang and Chen, Zhenpeng and Liu, Xuanzhe and Ma, Yun},
  journal={ACM Transactions on Software Engineering and Methodology},
  pages ={accepted to appear},
  year={2025}
}

@article{xiao2025bias,
  title={Bias in large {AI} models for medicine and healthcare: Survey and challenges},
  author={Xiao, Ying and Chen, Zhenpeng and Huang, Jen-tse and Chen, Wenting and Liu, Yepang and Li, Kezhi and Mousavi, Mohammad Reza and Dobson, Richard and Zhang, Jie M},
  year={2025},
  publisher={Preprints}
}

@article{hassan2024rethinking,
  title={Rethinking software engineering in the foundation model era: From task-driven AI copilots to goal-driven AI pair programmers},
  author={Hassan, Ahmed E and Oliva, Gustavo A and Lin, Dayi and Chen, Boyuan and Zhen Ming Jiang},
  journal={arXiv preprint arXiv:2404.10225},
  year={2024}
}

@inproceedings{AlshahwanHHMSW24,
  author       = {Nadia Alshahwan and
                  Mark Harman and
                  Inna Harper and
                  Alexandru Marginean and
                  Shubho Sengupta and
                  Eddy Wang},
  title        = {Assured LLM-based software engineering},
  booktitle    = {Proceedings of the 2nd {IEEE/ACM} International Workshop on Interpretability, Robustness,
                  and Benchmarking in Neural Software Engineering, InteNSE@ICSE 2024},
  pages        = {7--12},
  year         = {2024}
}

@article{harman2001search,
  title={Search-based software engineering},
  author={Harman, Mark and Jones, Bryan F},
  journal={Information and software Technology},
  volume={43},
  number={14},
  pages={833--839},
  year={2001}
}

@article{sahoo2024systematic,
  title={A systematic survey of prompt engineering in large language models: Techniques and applications},
  author={Sahoo, Pranab and Singh, Ayush Kumar and Saha, Sriparna and Jain, Vinija and Mondal, Samrat and Chadha, Aman},
  journal={arXiv preprint arXiv:2402.07927},
  year={2024}
}

@inproceedings{tafreshipour2024prompting,
  author       = {Tafreshipour, Mahan and Imani, Aaron and Huang, Eric and Almeida, Eduardo and Zimmermann, Thomas and Ahmed, Iftekhar},
  title        = {Prompting in the wild: An empirical study of prompt evolution in software repositories},
  booktitle    = {Proceedings of the 22nd {IEEE/ACM} International Conference on Mining Software Repositories,
                  {MSR} 2025},
pages        = {686--698},
  year         = {2025}
}

@article{guo2026eet,
  title={EET: Experience-driven early termination for cost-efficient software engineering agents},
  author={Guo, Yaoqi and Xiao, Ying and Zhang, Jie M and Harman, Mark and Lou, Yiling and Liu, Yang and Chen, Zhenpeng},
  journal={arXiv preprint arXiv:2601.05777},
  year={2026}
}

@inproceedings{yang2025bamas,
  title={BAMAS: Structuring budget-aware multi-agent systems},
  author={Yang, Liming and Luo, Junyu and Liu, Xuanzhe and Lou, Yiling and Chen, Zhenpeng},
  booktitle    = {Proceedings of the 40th Annual AAAI Conference on Artificial Intelligence, AAAI 2026},
  year={2026}
}

@inproceedings{LLMcain2025,
  title={Themes of building LLM-based applications for production: A practitioner's view},
  author={Mailach, Alina and Simon, Sebastian and Dorn, Johannes and Siegmund, Norbert},
  booktitle={Proceedings of the 2025 IEEE/ACM 4th International Conference on AI Engineering--Software Engineering for AI, CAIN 2025},
  pages={18--30},
  year={2025}
}

@inproceedings{debenedetti2024agentdojo,
  title={AgentDojo: A dynamic environment to evaluate prompt injection attacks and defenses for LLM agents},
  author={Debenedetti, Edoardo and Zhang, Jie and Balunovic, Mislav and Beurer-Kellner, Luca and Fischer, Marc and Tram{\`e}r, Florian},
  booktitle={Proceedings of Advances in Neural Information Processing Systems 38: Annual Conference
                  on Neural Information Processing Systems 2024, NeurIPS 2024},
  year={2024}
}

@article{PromptsArePrograms,
author = {Liang, Jenny T. and Lin, Melissa and Rao, Nikitha and Myers, Brad A.},
title = {Prompts are programs too! Understanding how developers build software containing prompts},
year = {2025},
volume = {2},
number = {FSE},
journal = {Proceedings of the ACM on Software Engineering},
pages={1591--1614}
}

@inproceedings{chenetal2024humans,
    title = "Humans or {LLM}s as the judge? A study on judgement bias",
    author = "Chen, Guiming Hardy  and
      Chen, Shunian  and
      Liu, Ziche  and
      Jiang, Feng  and
      Wang, Benyou",
    booktitle = "Proceedings of the 2024 Conference on Empirical Methods in Natural Language Processing, EMNLP 2024",
    year = "2024",
    pages = "8301--8327",
}

@inproceedings{luo2014empirical,
  title={An empirical analysis of flaky tests},
  author={Luo, Qingzhou and Hariri, Farah and Eloussi, Lamyaa and Marinov, Darko},
  booktitle={Proceedings of the 22nd ACM SIGSOFT international symposium on foundations of software engineering, FSE 2014},
  pages={643--653},
  year={2014}
}

@article{ouyang2024empirical,
  title={An empirical study of the non-determinism of ChatGPT in code generation},
  author={Ouyang, Shuyin and Zhang, Jie M and Harman, Mark and Wang, Meng},
  journal={ACM Transactions on Software Engineering and Methodology},
  volume       = {34},
  number       = {2},
  pages        = {42:1--42:28},
  year         = {2025}
}

@inproceedings{gligoric2013comparing,
  title={Comparing non-adequate test suites using coverage criteria},
  author={Gligoric, Milos and Groce, Alex and Zhang, Chaoqiang and Sharma, Rohan and Alipour, Mohammad Amin and Marinov, Darko},
  booktitle={Proceedings of the 2013 International Symposium on Software Testing and Analysis, ISSTA 2013},
  pages={302--313},
  year={2013}
}

@article{csurChenKLPTTZ18,
  author       = {Tsong Yueh Chen and
                  Fei{-}Ching Kuo and
                  Huai Liu and
                  Pak{-}Lok Poon and
                  Dave Towey and
                  T. H. Tse and
                  Zhi Quan Zhou},
  title        = {Metamorphic testing: {A} review of challenges and opportunities},
  journal      = {ACM Computing Surveys},
  volume       = {51},
  number       = {1},
  pages        = {4:1--4:27},
  year         = {2018}
}

@inproceedings{HassanLRGC00TOL24,
  author       = {Ahmed E. Hassan and
                  Dayi Lin and
                  Gopi Krishnan Rajbahadur and
                  Keheliya Gallaba and
                  Filipe Roseiro C{\^{o}}go and
                  Boyuan Chen and
                  Haoxiang Zhang and
                  Kishanthan Thangarajah and
                  Gustavo Ansaldi Oliva and
                  Jiahuei (Justina) Lin and
                  Wali Mohammad Abdullah and
                  Zhen Ming (Jack) Jiang},
  title        = {Rethinking software engineering in the era of foundation models: {A}
                  curated catalogue of challenges in the development of trustworthy
                  FMware},
  booktitle    = {Companion Proceedings of the 32nd {ACM} International Conference on
                  the Foundations of Software Engineering, {FSE} 2024},
  pages        = {294--305},
  year         = {2024}
}

@inproceedings{nahar2024beyond,
  title={Beyond the comfort zone: Emerging solutions to overcome challenges in integrating LLMs into software products},
  author={Nahar, Nadia and K{\"a}stner, Christian and Butler, Jenna and Parnin, Chris and Zimmermann, Thomas and Bird, Christian},
  booktitle={Proceedings of the 47th {IEEE/ACM} International Conference on Software Engineering, Software Engineering in Practice Track,
                  {ICSE-SEIP} 2025},
  pages        = {516--527},
  year={2025}
}

@inproceedings{huang2024apathetic,
  title={Apathetic or empathetic? Evaluating LLMs' emotional alignments with humans},
  author={Huang, Jen-tse and Lam, Man Ho and Li, Eric John and Ren, Shujie and Wang, Wenxuan and Jiao, Wenxiang and Tu, Zhaopeng and Lyu, Michael},
  booktitle={Proceedings of Advances in Neural Information Processing Systems 38: Annual Conference
                  on Neural Information Processing Systems 2024, NeurIPS 2024},
  year={2024}
}

@inproceedings{giallanza2024context,
  title={Context-sensitive semantic reasoning in large language models},
  author={Giallanza, Tyler and Campbell, Declan Iain},
  booktitle={ICLR 2024 Workshop on Representational Alignment},
  year={2024}
}

@inproceedings{tseng2024two,
  author       = {Yu{-}Min Tseng and
                  Yu{-}Chao Huang and
                  Teng{-}Yun Hsiao and
                  Wei{-}Lin Chen and
                  Chao{-}Wei Huang and
                  Yu Meng and
                  Yun{-}Nung Chen},
  title        = {Two tales of persona in {LLMs}: {A} survey of role-playing and personalization},
  booktitle    = {Findings of the Association for Computational Linguistics: {EMNLP}
                  2024},
  pages        = {16612--16631},
  year         = {2024}
}

@inproceedings{WanWHGBL23,
  author       = {Yuxuan Wan and
                  Wenxuan Wang and
                  Pinjia He and
                  Jiazhen Gu and
                  Haonan Bai and
                  Michael R. Lyu},
  title        = {BiasAsker: Measuring the bias in conversational {AI} system},
  booktitle    = {Proceedings of the 31st {ACM} Joint European Software Engineering
                  Conference and Symposium on the Foundations of Software Engineering,
                  {ESEC/FSE} 2023},
  pages        = {515--527},
  year         = {2023}
}

@article{xinyuerole,
  author       = {Xinyue Li and
                  Zhenpeng Chen and
                  Jie M. Zhang and
                  Yiling Lou and
                  Tianlin Li and
                  Weisong Sun and
                  Yang Liu and
                  Xuanzhe Liu},
  title        = {Benchmarking bias in large language models during role-playing},
  journal      = {CoRR},
  volume       = {abs/2411.00585},
  year         = {2024}
}

@inproceedings{diaz2024large,
  title={Large language model operations (LLMOps): Definition, challenges, and lifecycle management},
  author={Diaz-De-Arcaya, Josu and L{\'o}pez-De-Armentia, Juan and Mi{\~n}{\'o}n, Ra{\'u}l and Ojanguren, Iker Lasa and Torre-Bastida, Ana I},
  booktitle={Proceedings of the 9th International Conference on Smart and Sustainable Technologies, SpliTech 2024},
  pages={1--4},
  year={2024}
}

@article{TantithamthavornPKC25,
  author       = {Chakkrit Kla Tantithamthavorn and
                  Fabio Palomba and
                  Foutse Khomh and
                  Joselito Joey Chua},
  title        = {MLOps, LLMOps, FMOps, and beyond},
  journal      = {{IEEE} Software},
  volume       = {42},
  number       = {1},
  pages        = {26--32},
  year         = {2025}
}

@inproceedings{KhattabSMZSVHSJ24,
  author       = {Omar Khattab and
                  Arnav Singhvi and
                  Paridhi Maheshwari and
                  Zhiyuan Zhang and
                  Keshav Santhanam and
                  Sri Vardhamanan and
                  Saiful Haq and
                  Ashutosh Sharma and
                  Thomas T. Joshi and
                  Hanna Moazam and
                  Heather Miller and
                  Matei Zaharia and
                  Christopher Potts},
  title        = {DSPy: Compiling declarative language model calls into state-of-the-art
                  pipelines},
  booktitle    = {Proceedings of the Twelfth International Conference on Learning Representations,
                  {ICLR} 2024},
  year         = {2024}
}

@book{Antonio,
  title={Agentic design patterns: A hands-on guide to building intelligent systems},
  author={ Antonio Gullí},
  year={2025},
  publisher={Springer}
}

@inproceedings{WSLCNCZ23,
  author       = {Xuezhi Wang and
                  Jason Wei and
                  Dale Schuurmans and
                  Quoc V. Le and
                  Ed H. Chi and
                  Sharan Narang and
                  Aakanksha Chowdhery and
                  Denny Zhou},
  title        = {Self-consistency improves chain of thought reasoning in language models},
  booktitle    = {Proceedings of the Eleventh International Conference on Learning Representations,
                  {ICLR} 2023},
  year         = {2023}
}

@inproceedings{YaoYZS00N23,
  author       = {Shunyu Yao and
                  Dian Yu and
                  Jeffrey Zhao and
                  Izhak Shafran and
                  Tom Griffiths and
                  Yuan Cao and
                  Karthik Narasimhan},
  title        = {Tree of thoughts: Deliberate problem solving with large language models},
  booktitle    = {Proceedings of Advances in Neural Information Processing Systems 36: Annual Conference
                  on Neural Information Processing Systems 2023, NeurIPS 2023},
  year         = {2023}
}

@book{oshin2025learning,
  title={Learning LangChain: Building AI and LLM applications with LangChain and LangGraph},
  author={Oshin, Mayo and Campos, Nuno},
  year={2025},
  publisher={O'Reilly Media, Inc.}
}

@inproceedings{XingLCHZSL25,
  author       = {Zhenchang Xing and
                  Yang Liu and
                  Zhuo Cheng and
                  Qing Huang and
                  Dehai Zhao and
                  Daniel Sun and
                  Chenhua Liu},
  title        = {When prompt engineering meets software engineering: {CNL-P} as natural
                  and robust ``APIs''' for human-AI interaction},
  booktitle    = {Proceedings of the Thirteenth International Conference on Learning Representations,
                  {ICLR} 2025},
  year         = {2025}
}

@inproceedings{yaoqipersonality,
  author       = {Yaoqi Guo and
                  Zhenpeng Chen and
                  Jie M. Zhang and
                  Yang Liu and
                  Yun Ma},
  title        = {Personality-guided code generation using large language models},
  booktitle    = {Proceedings of the 63rd Annual Meeting of the Association for Computational
                  Linguistics (Volume 1: Long Papers), {ACL} 2025},
  pages        = {1068--1080},
  year         = {2025}
}

@inproceedings{li2023large,
  title={Large language models in finance: A survey},
  author={Li, Yinheng and Wang, Shaofei and Ding, Han and Chen, Hang},
  booktitle={Proceedings of the fourth ACM International Conference on AI in Finance, ICAIF 2023},
  pages={374--382},
  year={2023}
}

@article{wang2024large,
  title={Large language models for education: A survey and outlook},
  author={Wang, Shen and Xu, Tianlong and Li, Hang and Zhang, Chaoli and Liang, Joleen and Tang, Jiliang and Yu, Philip S and Wen, Qingsong},
  journal={arXiv preprint arXiv:2403.18105},
  year={2024}
}

@article{touvron2023llama,
  title={Llama: Open and efficient foundation language models},
  author={Touvron, Hugo and Lavril, Thibaut and Izacard, Gautier and Martinet, Xavier and Lachaux, Marie-Anne and Lacroix, Timoth{\'e}e and Rozi{\`e}re, Baptiste and Goyal, Naman and Hambro, Eric and Azhar, Faisal and Aurelien Rodriguez and Armand Joulin and Edouard Grave and Guillaume Lample},
  journal={arXiv preprint arXiv:2302.13971},
  year={2023}
}

@article{achiam2023gpt,
  title={GPT-4 technical report},
  author={Achiam, Josh and Adler, Steven and Agarwal, Sandhini and Ahmad, Lama and Akkaya, Ilge and Aleman, Florencia Leoni and Almeida, Diogo and Altenschmidt, Janko and Altman, Sam and Anadkat, Shyamal and others},
  journal={arXiv preprint arXiv:2303.08774},
  year={2023}
}

\end{document}